\documentclass[a4paper,12pt]{article}
\usepackage[latin1]{inputenc}
\usepackage[T1]{fontenc}
\usepackage{amsmath, amsthm, amssymb}
\usepackage{array}

\usepackage[final]{graphicx}   
\graphicspath{{tikz/}}

\newtheorem{definition}{Definition}[section]
\newtheorem{theorem}[definition]{Theorem}

\newtheorem{remark}[definition]{Remark}

\newtheorem{proposition}[definition]{Proposition}
\title{A decoding algorithm for 2D convolutional codes over the erasure channel}
\author{Julia Lieb, \thanks{
Julia Lieb is at the Institute of Mathematics, University of Zurich, Switzerland,  e-mail: julia.lieb@math.uzh.ch}\\ 
and Raquel Pinto\thanks{
Raquel Pinto is at the Department of Mathematics, University of Aveiro, Portugal e-mail: raquel@ua.pt
}}
\begin{document}

\maketitle

\begin{abstract}
Two-dimensional (2D) convolutional codes are a generalization of (1D) convolutional codes, which are very appropriate for transmission over an erasure channel. In this paper, we present a decoding algorithm for 2D convolutional codes over this kind of channel by reducing the decoding process to several decoding steps with 1D convolutional codes. Moreover, we provide constructions of 2D convolutional codes that are specially taylored to our decoding algorithm.
\end{abstract}

\pagenumbering{arabic}
\pagestyle{plain}

\section{Introduction}

An erasure channel is a communication channel where the receiver
knows if a received symbol is correct since symbols either arrive correctly or are erased. It is commonly used for multimedia traffic like the Internet.
When transmitting over such a channel, convolutional codes are very suitable. This is due to their ability of sliding along the sequence of received symbols in windows of variable size adapted to the location and frequency of the erasures.\\

Multidimensional convolutional codes generalize one-dimensional (1D) convolutional codes in a natural way. They can be applied for
transmission of multidimensional data, such as pictures or videos (2D) or animations (3D). Up to now there is not so much known about multidimensional convolutional codes. Two-dimensional (2D) convolutional codes were introduced in 1994 by Fornasini and Valcher \cite{f}, while multidimensional convolutional codes in general were first studied in 1998 by Weiner \cite{w}. There are some more research works dealing with 2D convolutional codes \cite{jdcr,dcr,dec} but decoding of convolutional codes in general is already hard in the one-dimensional case.
%
%
However, for the erasure channel, the decoding of 1D convolutional codes can be reduced to linear algebra operations, see \cite{vp}. In this paper, we reduce the decoding of 2D convolutional codes to the decoding of several 1D convolutional codes. This allows us to transfer results from the one-dimensional to the two-dimensional setting. However, additionally, we have to take care about a few details concerning the relation between the several 1D convolutional codes.\\

Our considerations in this paper have some similarity to the recovering of burst of erasures on horizonal/vertical lines in \cite{dec}. But in contrast to that paper, which only uses the parity-check matrix of the convolutional code, we also use the generator matrix of the 2D convolutional code, which enables us to considere the decoding in the corresponding 1D convolutional codes not isolated but to employ relations between the different steps in the decoding process. Moreover, to the best of our knowledge, we present the first complete decoding algorithm for 2D convolutional codes. Furthermore, we present constructions of 2D convolutional codes targeted to our decoding algorithm.\\

The paper is organized as follows: In Section 2.1, we summarize some basic results on 1D convolutional codes and also present some new results such as an erasure decoding algorithm for 1D convolutional codes and describe how it is possible to use this algorithm also in the case that the corresponding code is catastrophic. In Section 2.2., we introduce 2D convolutional codes. In Section 3., we present the different parts of our main decoding algorithm and illustrate it with examples. In Section 4, we provide constructions for different code rates that are targeted to our decoding algorithm. In Section 5, we evaluate the erasure correcting capability of our decoding algorithm. Finally, in Section 6, we conclude with some remarks.

\section{Convolutional codes}

In this section, we present the necessary background about 1D and 2D convolutional codes that is important for our decoding algorithm for 2D convolutional codes in Section \ref{dec} and the taylored construction in Section \ref{con}. For more details about the theory of convolutional codes, see \cite{ch}. Moreover, we present a decoding algorithm for 1D convolutional codes that will be a building block for our 2D decoding algorithm.

\subsection{1D Convolutional codes}

\begin{definition}
An $(n,k)$ one-dimensional (1D) \textbf{convolutional code} $\cal C$ is an $\mathbb F[z]$-submodule of $\mathbb F[z]^n$ of rank $k$.
A full column rank polynomial matrix $G(z) \in \mathbb F[z]^{n \times k}$ whose columns constitute a basis of $\cal C$ is called a \textbf{generator matrix} of $\cal C$ and we have
that
\begin{eqnarray*}
  {\cal C} &=&Im_{\mathbb F[z]}G(z) \\
   &=& \{G(z)u(z) \, | \, u(z) \in \mathbb F[z]^k\}.
\end{eqnarray*}
\end{definition}

An $\mathbb F[z]$-submodule of $\mathbb F[z]^n$ admits many bases and therefore an $(n,k)$ convolutional code $\cal C$ has many generator matrices. Generator matrices of the same code are said to be equivalent and differ by right multiplication with a unimodular matrix (a $k \times k$ invertible polynomial matrix with polynomial inverse).\\

Therefore, the full size minors of two generator matrices of a convolutional code differ by a nonzero constant, and consequently the highest degree of the full size minors of all generator matrices of a convolutional code is the same and it is called the \textbf{degree} of the code. An $(n,k)$ convolutional code with degree $\delta$ is said to be an $(n,k,\delta)$ code.
\\

An important property of the generator matrices that reflects on the properties of the corresponding convolutional code is primeness. A full column rank polynomial matrix $G(z)\in \mathbb F[z]^{n \times k}$ is \textbf{right prime} if
$$
G(z)= \bar G(z) X(z),
$$
for $\bar G(z)\in \mathbb F[z]^{n \times k}$ and $X(z) \in \mathbb F[z]^{k \times k}$ implies that $X(z)$ must be unimodular. Therefore if $G(z)$ is a right prime generator matrix of an $(n,k)$ convolutional code then all the generator matrices of the code are also right prime. In the same way if $G(z)$ is not right prime then all its equivalent generator matrices are not right prime. This means that right primeness of the generator matrices of a code is a property of the code and we say that a convolutional code is \textbf{noncatastrophic} if all its generator matrices are right prime. Noncatastrophic convolutional codes are the ones that admit a kernel representation as stated in the next theorem.

\begin{theorem}\cite{y}\label{H}
Let $\cal C$ be an $(n,k)$ convolutional code. Then $\cal C$ is noncatastrophic if and only if there exists a full row rank matrix $H(z) \in \mathbb F[z]^{(n-k) \times n}$ such that
\begin{eqnarray*}
  {\cal C} &=& Ker_{\mathbb F[z]}H(z) \\
   &=& \{v(z) \in \mathbb F[z]^n \, | \, H(z)v(z)=0\}.
\end{eqnarray*}
\end{theorem}

Let $\cal C$ be a noncatastrophic convolutional code. A left prime matrix $H(z) \in \mathbb F[z]^{(n-k) \times n}$ such that ${\cal C} = Ker_{\mathbb F[z]}H(z)$ is called a \textbf{parity-check matrix} of $\cal C$, and from Theorem \ref{H} we have that for $v(z) \in \mathbb F[z]^{n}$,
$$
v(z) \in {\cal C} \Leftrightarrow H(z)v(z)=0.
$$

Parity-check matrices of a convolutional code are very important for information transmission over an erasure channel. In these type of channels, a symbol arrives correctly or it does not arrive and is considered an erasure. Thus a word that is received after channel transmission is a codeword of the code with some symbols missing. In \cite{vp} the authors present a decoding algorithm for noncatastrophic convolutional codes which uses the parity-check matrix of the code, as explained next.\\

Assume that $v(z)=\sum_{i\in\mathbb N_0}v_iz^{i}\in\mathcal{C}$ is sent and the coefficients $v_0,\hdots,v_{t-1}$ arrive (correctly) for some $t\in\mathbb N_0$ and at least one component of the vector $v_t$ is erased. If $H(z)=\sum_{i=0}^{\nu}H_iz^{i}$ is a parity-check matrix of $\mathcal{C}$ then, for each $j\in\mathbb N_0$ and
\begin{align}
\mathfrak{H}_j:=\left(
  \begin{array}{ccccc} H_{\nu} & \cdots & H_0 & &  0\\
     & \ddots & & \ddots & \\
    0 & & H_{\nu} & \cdots & H_0 \\
    \end{array}
\right)\in \mathbb F^{(j+1)(n-k)\times (\nu+j+1)n},
\end{align}
one has $\mathfrak{H}_j [v_{t-\nu}, \hdots ,v_{t+j}]=\mathbf{0}$, where $v_i=0$ for $i\notin\{0,\hdots,\deg(v)\}$. 

Denote by $H_j^c$ the matrix consisting of the last $(j+1)n$ columns of $\mathfrak{H}_j$, by $v^{(e)}_i$ the erased components of $v_i$, and by $H_j^{c,(e)}$ the corresponding columns of $H^c_j$. Then, recovering $[v^{(e)}_t, \hdots ,v^{(e)}_{t+j}]$ is equivalent to solving a system of linear equations of the form 
\begin{align}\label{system}
H_j^{c,(e)}[v^{(e)}_t, \hdots ,v^{(e)}_{t+j}]=b,
\end{align}
where $b\in\mathbb F^{(j+1)(n-k)}$ is known.
%
The erasures $[v^{(e)}_t, \hdots ,v^{(e)}_{t+j}]$ are recovered if and only if the system has a unique solution, i.e. if and only if $H_j^{c,(e)}$ has full column rank.\\

The following theorem characterizes the capability of erasure correction of a noncatastrophic convolutional code in terms of its parity-check matrices. It is a consequence of Theorem 3.1. of \cite{vp}. This capability is directly connected with the notion of column distance of the code as defined next.

\begin{definition}
The \textbf{Hamming weight} $wt(v)$ of $v\in\mathbb F^n$ is defined as the number of its nonzero components.\\
For $v(z)\in\mathbb F[z]^n$ with $\deg(v(z))=\gamma$, write $v(z)=v_0+\cdots+v_{\gamma}z^{\gamma}$ with $v_t\in\mathbb F^n$ for $t=0,\hdots,\gamma$ and set $v_t=0\in\mathbb F^n$ for $t\geq\gamma+1$. Then, for $j\in\mathbb N_0$, the \textbf{j-th column distance} of a convolutional code $\mathcal{C}$ is defined as
$$d_j^c(\mathcal{C}):=\min_{v(z)\in\mathcal{C}}\left\{\sum_{t=0}^j wt(v_t)\ |\ v_0\neq 0\right\}.$$
Moreover, $d_{free}(\mathcal{C}):=\min_{v(z)\in\mathcal{C}}\left\{\sum_{t=0}^{\deg(v(z))} wt(v_t)\ |\ v(z)\not\equiv 0\right\}$ is called the \textbf{free distance} of $\mathcal{C}$. It holds $d_0^c\leq d_1^c\leq\cdots\leq d_{free}$ and $d_{free}(\mathcal{C})=\lim_{j\rightarrow\infty}d_j^c(\mathcal{C})$.
\end{definition}

\begin{theorem}\cite{strongly}\label{pattern}
  Let $\cal C$ be an $(n,k)$ noncatastrophic convolutional code and $H(z)$ a parity-check matrix. Then the following are equivalent:
  \begin{enumerate}
  \item the $j$-th column distance of $\cal C$ is $d$;
    \item none of the first $n$ columns of $H^c_j$ is contained in the span of any other $d-2$ columns and one of the first $n$ columns of $H^c_j$ is in the span of some other $d-1$ columns of the matrix;
    \item if in a sliding window of length $j+1$ at most $d-1$ erasures occur and the preceding symbols are correct, then we can completely recover the first $n$ symbols in the sliding window.
  \end{enumerate}
  \end{theorem}

The preceding theorem leads to the following decoding algorithm for 1D convolutional codes. Assume that we receive $\hat{v}(z)=\sum_{0\leq i\leq s}\hat{v}_{i}z^{i}$, where each component of the vectors $\hat{v}_{i}$ is either identical with the corresponding component of the sent vector $v_{i}$ or it is erased and write the symbol $\ast$ for each erased component. Moreover, denote by $\epsilon_i$ the number of erased components of $\hat{v}_i$ and by $d^c_j$ the $j$-th column distance of the convolutional code.\\


\textbf{Algorithm 1}

\textbf{1}: Set $i=0$.\\
\textbf{2}: Set $j=0$.\\
\textbf{3}: If $\epsilon_i+\hdots+\epsilon_{i+j}\leq d^c_j-1$, go to 5, otherwise go to 4.\\
\textbf{4}: If $d^c_j=d_{free}$, we cannot recover the erasures in $v_i$, otherwise set $j=j+1$ and go to 3.\\
\textbf{5}: Recover the erasures in $\hat{v}_i$ using Theorem \ref{pattern} and solving the system of linear equations \eqref{system}. Replace the $\ast$ symbols with the correct symbols.\\
\textbf{6}: If $i=s$, the decoding is finished, otherwise set $i=i+1$ and go to 2.

\begin{remark}
A decoding algorithm for 1D convolutional codes over the erasure channel can also be found in \cite{j}. There the focus is more on the decoding with low delay and how to restart the decoding if part of the sequence has to be declared as lost. Here we concentrate on full recovery and formulate it in terms of the column distances of the code.
\end{remark}

The following theorem is an immediate consequence of Theorem \ref{pattern} and describes which erasure patterns can be corrected with Algorithm 1.

\begin{theorem}
Let $\mathcal{C}$ be a convolutional code with column distances $d_j^c$ and assume that we receive $\hat{v}(z)=\sum_{0\leq i\leq s}\hat{v}_{i}z^{i}$, where the number of erased components of $\hat{v}_i$ is denoted by $\epsilon_i$.
If for each $0\leq i\leq s$, there exists a $j\in\mathbb N_0$ such that $\epsilon_i+\cdots+\epsilon_{i+j}\leq d_j^c-1$, then all erasures can be recovered.
\end{theorem}

The erasure correcting capability of a convolutional code increases with its column distances, which are upper bounded as the following theorem shows.

\begin{theorem}\cite{strongly}\label{ub}
Let $\mathcal{C}$ be an $(n,k,\delta)$ convolutional code. Then, it holds:
$$d_j^c (\mathcal{C}) \leq (n-k)(j + 1) + 1\ \ \text{for}\ \ j\in\mathbb N_0.$$
\end{theorem}

The column distances of a convolutional code could reach this upper bound only up to $j=L:=\left\lfloor\frac{\delta}{k}\right\rfloor+\left\lfloor\frac{\delta}{n-k}\right\rfloor$.

\begin{definition}\cite{mdp}
A convolutional code $\mathcal{C}$ of rate $k/n$ and degree $\delta$ has
  \textbf{maximum distance profile (MDP)} if
$$d_j^c(\mathcal{C})=(n-k)(j+1)+1\quad \text{for}\ j=0,\hdots,L:=\left\lfloor\frac{\delta}{k}\right\rfloor+\left\lfloor\frac{\delta}{n-k}\right\rfloor$$
\end{definition}

According to \cite{strongly}, it is sufficient to have equality for $j=L$ in Theorem  \ref{ub} to get an MDP convolutional code. Moreover, one has the following theorem to check if a convolutional code is MDP.


\begin{theorem}\label{cd}\cite{strongly}
Let $\mathcal{C}$ have generator matrix $G(z)=\sum_{i=0}^{\mu}G_iz^i\in\mathbb F[z]^{n\times k}$ and parity-check $H(z)=\sum_{i=0}^{\nu}H_iz^i\in\mathbb F[z]^{n-k\times n}$. The following statements are equivalent:
\begin{itemize}
\item[(i)] $d_j^c (\mathcal{C})=(n-k)(j + 1) + 1$ 
\item[(ii)] $G^c_j:=\left[\begin{array}{ccc} G_0 & & 0\\ \vdots & \ddots &  \\ G_j & \hdots & G_0 \end{array}\right]$ where $G_i\equiv 0$ for $i>\mu$ has the property that every full size minor that is not trivially zero, i.e. zero for all choices of $G_1,\hdots,G_j$, is nonzero.
\item[(iii)] $H_j^c:=\left[\begin{array}{ccc} H_0 & & 0\\ \vdots & \ddots &  \\ H_j & \hdots & H_0 \end{array}\right]$ with $H_i\equiv 0$ for $i>\nu$ has the property that every full size minor that is not trivially zero
is nonzero.
\end{itemize}
\end{theorem}

The decoding properties of an MDP convolutional code should be presented in the following.

\begin{proposition}\cite{vp}\label{mdpp}\ \\
If for an $(n,k,\delta)$ MDP convolutional code $\mathcal{C}$, in any sliding window of length at most $(L+1)n$ at most $(L+1)(n-k)$ erasures occur, then full error correction from left to right is possible.
\end{proposition}


To conclude this chapter, in the following, we will explain how to decode a convolutional code over the erasure channel, even if the code is catastrophic and hence admits no parity-check matrix.\\

Note that if $\cal C$ and $\bar{\cal C}$ are two noncatastrophic convolutional codes such that $\cal C \subset \bar{\cal C}$, then the $j$-th column distance of $\cal C$ is  greater than the $j$-th column distance of $\bar{\cal C}$ and therefore, by Theorem \ref{H}, $\cal C$ has better erasure correction capability than $\bar{\cal C}$. \\

As stated in Theorem \ref{H}, if $\cal C$ is not a noncatastrophic convolutional , it does not admit a parity-check matrix. However there is an $(n-k) \times n$ left prime matrix $H(z)$ such that ${\cal C} \subset Ker_{\mathbb F[z]} H(z)$. In fact, if $G(z) \in \mathbb F[z]^{n \times k}$ is an encoder of $\cal C$, then there exists a right prime matrix $\tilde G(z) \in \mathbb F[z]^{n \times k}$ such that $G(z)=\tilde G(z) X(z)$ for some nonsingular matrix $X(z) \in \mathbb F[z]^{k \times k}$. Then $\tilde {\cal C}=Im_{\mathbb F[z]} \tilde G(z)$ is the smallest noncatastrophic convolutional code such that ${\cal C} \subset \tilde {\cal C}$. Consequently, if $H(z) \in \mathbb F[z]^{(n-k) \times n}$ is a parity-check matrix of $\tilde {\cal C}$ it follows that ${\cal C} \subset Ker_{\mathbb F[z]} H(z)$, and we have that
\begin{equation}\label{cont}
v(z) \in {\cal C} \Rightarrow H(z)v(z)=0,
\end{equation}
for any $v(z) \in \mathbb F[z]^n$. Thus we can use the matrix $H(z)$ for decoding over the erasure channel, since all the received words $v(z)$ are codewords  of $\cal C$ (with some erasures in it) and therefore must satisfy the equation $H(z)v(z)=0$ .\\

Finally, let us assume that ${\cal C}_0$ is an $(n,k)$ convolutional code with encoder $G_0(z)$ and let $H_0(z) \in\mathbb F[z]^{(n-k) \times n}$ be a left prime matrix such that $\tilde {\cal C}_0 = Ker_{\mathbb F[z]} H_0(z)$ is the smallest noncatastrophic convolutional code that contains ${\cal C}_0$. Let $G_1(z) \in \mathbb F[z]^{n \times k_1}$ be full column rank with $k_1 \in \mathbb N$ such that $k+k_1 <n$, and consider the $(n,k+k_1)$ convolutional code ${\cal C}_1$ with encoder $[G_0(z) \; G_1(z)]$. Let $H_1(z) \in\mathbb F[z]^{(n-k-k_1) \times n}$ be a left prime matrix such that $\tilde {\cal C}_1 = Ker_{\mathbb F[z]} H_1(z)$ is the smallest noncatastrophic convolutional code that contains ${\cal C}_1$. Next we show that $\tilde {\cal C}_0 \subset \tilde {\cal C}_1$.

Let $\tilde G_0(z) \in \mathbb F[z]^{n \times k}$ and $\tilde G_1(z) \in \mathbb F[z]^{n \times (k+k_1)}$ be two encoders of $\tilde {\cal C}_0$ and $\tilde {\cal C}_1$, respectively. Then $G_0(z)=\tilde G_0(z) X_0(z)$ for some invertible matrix $X_0(z) \in \mathbb F[z]^{k \times k}$, and therefore
$$
[G_0(z) \; G_1(z)] = [\tilde G_0(z) \; G_1(z)] \left[\begin{array}{cc} X(z) & 0 \\ 0 & I_{k_1} \end{array} \right].
$$
We conclude that ${\cal C}_1 \subset Im_{\mathbb F[z]}[\tilde G_0(z) \; G_1(z)]$. Note that also\\ $\tilde {\cal C}_0 \subset Im_{\mathbb F[z]}[\tilde G_0(z) \; G_1(z)]$. On the other hand let $\hat G_1(z) \in \mathbb F[z]^{n \times {k + k_1}}$ be a right prime matrix and $X_1(z) \in \mathbb F[z]^{(k+k_1) \times {k + k_1}}$ an invertible matrix such that
$$
[\tilde G_0(z) \; G_1(z)]=\hat G_1(z) X_1(z).
$$
Then $Im_{\mathbb F[z]}[\tilde G_0(z) \; G_1(z)] \subset Im_{\mathbb F[z]} \hat G_1(z)$ and, moreover,
$$
[G_0(z) \; G_1(z)] = \hat G_1(z) X_1(z) \left[\begin{array}{cc} X(z) & 0 \\ 0 & I_{k_1} \end{array} \right],
$$
which means that $\tilde {\cal C}_1=Im_{\mathbb F[z]} \hat G_1(z)$ and consequently $\tilde {\cal C}_0 \subset \tilde {\cal C}_1$.

\subsection{2D convolutional codes}


In this section we briefly introduce two-dimensional (2D) convolutional codes. 

Since in this paper, we will present a decoding algorithm that breaks down the decoding of a 2D convolutional code to several decoding steps with 1D convolutional codes, there is not much background on 2D convolutional codes needed.

\begin{definition}
An $(n,k)$ \textbf{two-dimensional (2D) convolutional code} $\mathcal{C}$ is a free $\mathbb F[z_1,z_2]$-submodule of $\mathbb F[z_1,z_2]^n$ of rank $k$. A \textbf{generator matrix} of $\mathcal{C}$ is a full row rank matrix $G(z_1,z_2)$ whose rows constitute a basis of $\mathcal{C}$, i.e.
\begin{eqnarray*}
  {\cal C} &=&Im_{\mathbb F[z_1,z_2]}G(z_1,z_2) \\
   &=& \{G(z_1,z_2)u(z_1,z_2) \, | \, u(z_1,z_2) \in \mathbb F[z_1,z_2]^k\}.
\end{eqnarray*}
\end{definition}


2D convolutional codes have two notions of degree, the internal and the external degree but these are not needed for our purposes. The interested reader is referred to \cite{ch} for more background on 2D convolutional codes.
%
%
%
%

For our decoding algorithm, we consider the generator matrix
$$G(z_1,z_2)=\sum_{i,j}G_{ij}z_1^{i}z_2^{j}$$
with $\mu:=\deg(G)=\max\{i+j: G_{ij}\neq 0\}$
and write it in the form
\begin{align}\label{g2}
G(z_1,z_2)=\sum_{i=0}^{\mu_{1}}G_i^{(2)}(z_2)z_1^{i},\quad\text{where}\ G^{(2)}_{\mu_1}(z_2)\neq 0
\end{align}
with $\mu_1=\deg_{z_1}(G(z_1,z_2))$ and $G_i^{(2)}(z_2)=\sum_jG_{ij}z_2^{j}$, where $\deg(G_i^{(2)})\leq\deg(G)-i$.
We encode the message
$$u(z_1,z_2)=\sum_{i=0}^{m_1}u_i^{(2)}(z_2)z_1^{i}.$$


The resulting codeword has the form $$v(z_1,z_2)=\sum_{i=0}^{m_1+\mu_1}v_i^{(2)}(z_2)z_1^{i}$$
with
\begin{align}\label{v2}
v_i^{(2)}(z_2)=\sum_{l+k=i}G_l^{(2)}(z_2)u_k^{(2)}(z_2).
\end{align}
Here, we set $G_l^{(2)}(z_2)\equiv 0$ if $l>\mu_1$ and $u_k^{(2)}(z_2)\equiv 0$ if $k>m_1$.

Consequently, successful decoding is equal to retrieving the polynomial vector $u^{(2)}(z_2):=[u_0^{(2)}(z_2)^{\top},\hdots,u_{m_1}^{(2)}(z_2)^{\top}]^{\top}$.\\


Alternatively, one could write
$$G(z_1,z_2)=\sum_{j=0}^{\mu_{2}}G_j^{(1)}(z_1)z_2^{i},\quad\text{where}\ G^{(1)}_{\mu_2}(z_1)\neq 0$$
with $\mu_2=\deg_{z_2}(G(z_1,z_2))$ and $G_j^{(1)}(z_1)=\sum_iG_{ij}z_1^{i}$, where $\deg(G_j^{(1)})\leq\deg(G)-j$,
$$u(z_1,z_2)=\sum_{j=0}^{m_2}u_j^{(1)}(z_1)z_2^{j}$$
and
$$v(z_1,z_2)=\sum_{j=0}^{m_2+\mu_2}v_j^{(1)}(z_1)z_2^{j}$$
with
\begin{align}\label{v1}
v_j^{(1)}(z_1)=\sum_{l+k=j}G_l^{(1)}(z_1)u_k^{(1)}(z_1).
\end{align}
Here, we set $G_l^{(1)}(z_1)\equiv 0$ if $l>\mu_2$ and $u_k^{(1)}(z_1)\equiv 0$ if $k>m_2$.

Consequently, successful decoding is also equal to knowing the polynomial vector $u^{(1)}(z_1):=[u_0^{(1)}(z_1)^{\top},\hdots,u^{(1)}_{m_2}(z_1)^{\top}]^{\top}$.\\


In the next two sections, we will present a decoding algorithms for 2D convolutional codes and then we will give constructions of 2D convolutional codes that have a good performance on this algorithm.

\section{Decoding algorithm for 2D convolutional codes}\label{dec}

In this section, we will present a decoding algorithm for 2D convolutional codes over the erasure channel. The main algorithm consists of several sub-algorithms that will be presented in the following subsections before the final algorithm will be presented. The basic idea is to use \eqref{v1} and \eqref{v2} to break down the whole decoding process to one-dimensional decoding with respect to $z_1$ or $z_2$.\\

Assume that we receive $\hat{v}(z_1,z_2)=\sum_{0\leq i\leq d_1, 0\leq j\leq d_2}\hat{v}_{ij}z_1^{i}z_2^{j}$, where each component of the vectors $\hat{v}_{ij}$ is either identical with the corresponding component of the sent vector $v_{ij}$ or it is erased and write the symbol $\ast$ for each erased component.
For the following algorithms we use the following equation where $\alpha,\beta\in\{1,2\}$ with $\alpha\neq \beta$.

\begin{align}\label{sliding}
\begin{pmatrix}v_0^{(\beta)}(z_{\beta})\\ v_1^{(\beta)}(z_{\beta})\\ \vdots \\ v^{(\beta)}_{d_{\alpha}}(z_{\beta}) \end{pmatrix}=
\left[
\begin{array}{ccccccccc}
G_0^{(\beta)}(z_{\beta}) \\
G_1^{(\beta)}(z_{\beta}) & G_0^{(\beta)}(z_{\beta})\\
\vdots  &  \ddots\\
G^{(\beta)}_{\mu_{\alpha}}(z_{\beta}) & \cdots & G^{(\beta)}_0(z_{\beta})\\
& \ddots & \ddots\\
& G^{(\beta)}_{\mu_{\alpha}}(z_{\beta}) & \cdots & G^{(\beta)}_0(z_{\beta})
\end{array}
\right]\begin{pmatrix}u^{(\beta)}_0(z_{\beta})\\ \vdots\\ u^{(\beta)}_{m_{\alpha}}(z_{\beta})\end{pmatrix}
\end{align}

Moreover, choose $e_{\beta}\in\mathbb N_0$ as large as possible such that $[G^{(\beta)}_{e_{\beta}}(z_{\beta}),\hdots,G^{(\beta)}_0(z_{\beta})]$ generates a 1D convolutional code, i.e. such that $k(e_{\beta}+1)<n$ and\\ $[G^{(\beta)}_{e_{\beta}}(z_{\beta}),\hdots,G^{(\beta)}_0(z_{\beta})]$ is full column rank.
For $s=0,\hdots, e_{\beta}$, compute $H^{(\beta)}_{s}(z_{\beta})$ with $H^{(\beta)}_{s}(z_{\beta})[G^{(\beta)}_s(z_{\beta}),\hdots, G^{(\beta)}_0(z_{\beta})]=0$, see \eqref{cont}. \\
If we already have recovered all erasures in $\hat{v}_0(z_{\beta}), \hdots, \hat{v}_f(z_{\beta})$ for some $f\in\mathbb N_0$, we know $u_0(z_{\beta}), \hdots, u_f(z_{\beta})$ by using the equation\\ $v_i(z_{\beta})=\sum_{m+l=i,\ l\geq 1}G^{(\beta)}_lu_m(z_{\beta})+G_0^{(\beta)}u_i(z_{\beta})$ for $i=0,\hdots,f$ and the fact that $G^{(\beta)}_0(z_{\beta})$ is injective. For $g\leq e_{\beta}+1$, we have\\ $v_{f+g}(z_{\beta})=\sum_{l+m=f+g,\ m>f}G^{(\beta)}_lu_m(z_{\beta})+\sum_{l+m=f+g,\ m\leq f}G^{(\beta)}_lu_m(z_{\beta})$ and can decode $\hat{v}_{f+g}(z_{\beta})-\sum_{l+m=f+g,\ m\leq f}G^{(\beta)}_lu_m(z_{\beta})$, which has the same erasure pattern as $\hat{v}_{f+g}(z_{\beta})$, in the code generated by $[G^{(\beta)}_{g-1},\hdots,G^{(\beta)}_0](z_{\beta})$ using the equation
\begin{align*}
&H^{(\beta)}_{g-1}(z_{\beta})\left(v^{(\beta)}_{f+g}(z_{\beta})-\sum_{l+m=f+g,\ m\leq f}G^{(\beta)}_l(z_{\beta})u_m(z_{\beta})\right)=\\
&=H^{(\beta)}_{g-1}(z_{\beta})[G^{(\beta)}_{g-1}(z_{\beta}),\hdots,G^{(\beta)}_0(z_{\beta})]\begin{pmatrix} u^{(\beta)}_{f+1}(z_{\beta})\\ \vdots\\ u^{(\beta)}_{f+g}(z_{\beta})\end{pmatrix}=0
\end{align*}
(see \eqref{sliding}) applying Algorithm 1. Then, we can use the injectivity of\\ $[G^{(\beta)}_{g-1}(z_{\beta}),\hdots, G^{(\beta)}_{0}(z_{\beta})]$ to get $u_{f+1}(z_{\beta}),\hdots,u_{f+g}(z_{\beta})$. This idea is used by the following decoding algorithms. Algorithm 2.1, Algorithm 2.2 and Algorithm 2.3 are parts of the main Algorithm 2 and should be described first.

\subsection{Full recovery of blocks in one direction}

In this subsection, we will describe the decoding with respect to $z_1$ and with respect to $z_2$ separately. In the last subsection of this section, when we formulate the main algorithm, we will describe how to combine the decoding in both directions.\\


\textbf{Case 1}: $\mu_{\alpha}> e_{\beta}$\\

\underline{Assume first that the degree $m_{\alpha}$ of $u$ with respect to $z_{\alpha}$ is not known}\\

\textbf{Decoding algorithm 2.1}\\
\textbf{1}: Set $b=0$.\\
\textbf{2}: Set $x_b=e_{\beta}+1$.\\
\textbf{3}: If
$\hat{v}^{(\beta)}_{\sum_{t=0}^{b}x_{t}-1}(z_{\beta})$
has an erasure pattern that can be (completely) recovered with the code generated by
$[G^{(\beta)}_{x_b-1}(z_{\beta}),\hdots,G^{(\beta)}_0(z_{\beta})]$, set $a_1=1$ and use Algorithm 1 to recover $v^{(\beta)}_{\sum_{t=0}^{b}x_{t}-1}(z_{\beta})$ and to get $u^{(\beta)}_{\sum_{t=0}^{b-1}x_t}(z_{\beta}), \hdots, u^{(\beta)}_{\sum_{t=0}^{b}x_t-1}(z_{\beta})$. Replace the corresponding $\ast$-symbols in $\hat{v}^{(\beta)}_{\sum_{t=0}^{b}x_{t}-1}(z_{\beta})$ with the recovered symbols and and go to 5.\\
If $\hat{v}^{(\beta)}_{\sum_{t=0}^{b}x_{t}-1}(z_{\beta})$
has an erasure pattern that can not be (completely) recovered, set $x_b\rightarrow x_b-1$ and go to Step 4.\\
\textbf{4}: If $x_b\neq 0$, go to 3, if $x_b=0$, go to 7.\\
\textbf{5}: $b\rightarrow b+1$\\
\textbf{6}: Go back to 2.\\
\textbf{7}: If $\sum_{t=0}^{b}x_t-1=d_{\alpha}$, stop the whole algorithm with successful recovery.\\
If $\sum_{t=0}^{b}x_t-1<d_{\alpha}$, go back to main algorithm.\\

As mentioned before the preceding algorithm is part of the main Algorithm 2. However, depending on the erasure pattern of the received word, Algorithm 2.1 could be sufficient for recovering all erasures, as the following theorem describes.

\begin{theorem}\ \\
Algorithm 2.1 is able to recover all erasures if one has an erasure pattern such that there are indices $0\leq j_0<\cdots<j_l=\mu_{\alpha}+m_{\alpha}$ with $j_0\leq e_{\beta}$ and $j_k-(j_{k-1}+1)\leq e_{\beta}$ for $k=1,\hdots,l$ such that $v^{(\beta)}_{j_0}(z_{\beta})$ could be decoded in $[G^{(\beta)}_{j_0},\hdots, G^{(\beta)}_0](z_{\beta})$ and $v^{(\beta)}_{j_k}(z_{\beta})$ could be decoded in $[G^{(\beta)}_{j_k-(j_{k-1}+1)},\hdots, G^{(\beta)}_0](z_{\beta})$ for $k=1,\hdots,l$.
\end{theorem}


In this case, i.e. $\mu_{\alpha}>e_{\beta}$, and if one does not know $m_{\alpha}$, one is not able to proceed with Algorithm 2.1 if one was able to recover $u^{(\beta)}_0(z_{\beta}),\hdots, u^{(\beta)}_{x-1}(z_{\beta})$ for some $x\in\{0,\hdots, m_{\alpha}\}$ but it is not possible to decode $v^{(\beta)}_x(z_{\beta}),\hdots,v^{(\beta)}_{x+e}(z_{\beta})$. \\

\underline{Assume now that the degree of $u$ with respect to $z_{\beta}$ is known}\\
One can proceed exactly as before but has the advantage that one knows that $u^{(\beta)}_f(z_{\beta})=0$ for $f>m_{\alpha}$. Thus, some of the $v^{(\beta)}_l(z_{\beta})$ can maybe be decoded in codes with lower rates, which makes it easier. If for example $m_{\alpha}=2$, i.e. $u_3^{(\beta)}(z_{\beta})=0$, then $v^{(\beta)}_3(z_{\beta})$ can be decoded in the code with generator matrix $[G^{(\beta)}_{2},\hdots, G^{(\beta)}_0](z_{\beta})$ instead of $[G^{(\beta)}_{3},\hdots, G^{(\beta)}_0](z_{\beta})$.\\
Moreover, if $m_{\alpha}$ is known, it is enough to recover $v^{(\beta)}_0(z_{\beta}),\hdots,v^{(\beta)}_{m_{\alpha}}(z_{\beta})$ to obtain $u(z_1,z_2)$ completely.
To give a simple example for this, assume that $v^{(\beta)}_0(z_{\beta}),\cdots,v^{(\beta)}_{m_{\alpha}}(z_{\beta})$ arrived completely and that $v^{(\beta)}_{m_{\alpha}+1}(z_{\beta}),\cdots,v^{(\beta)}_{m_{\alpha}+\mu_{\alpha}}(z_{\beta})$ are completely erased. In this case, clearly full recovery is possible if one knows $m_{\alpha}$ but if one does not know it, no erasures could be recovered.
\\
Furthermore, if there exists $\tilde{e}_{\beta}\in\mathbb N_0$ such that $[G^{(\beta)}_{\mu_{\alpha}},\hdots, G^{(\beta)}_{\mu_{\alpha}-\tilde{e}_{\beta}}](z_{\beta})$ is the generator matrix of a convolutional code (which is true for Construction 2 presented later in this paper), one can also reverse the whole decoding process and start to decode $v^{(\beta)}_{\mu_{\alpha}+m_{\alpha}-x_0}(z_{\beta})$ in the code with generator matrix $[G^{(\beta)}_{\mu_{\alpha}},\hdots, G^{(\beta)}_{\mu_{\alpha}-x_0}](z_{\beta})$ where $x_0\leq\tilde{e}_{\beta}$ is chosen maximal such that this is possible. Therefore, if one is stuck in one decoding direction, one could move to the other and try there.
In this way, more erasure patterns could be recovered as in the case that $m_{\alpha}$ is not known.\\

\textbf{Case 2}: $\mu_{\alpha}=e_{\beta}$\\


\underline{Assume first that the degree $m_{\alpha}$ of $u$ with respect to $z_{\alpha}$ is not known}\\


\textbf{Decoding algorithm 2.2}

\textbf{0}: Set $r=\chi_{-1}=0$.\\
\textbf{1}: Set $b=c=0$.\\
\textbf{2}: Set $x_b=e_{\beta}+1$.\\
\textbf{3}: If
$\hat{v}^{(\beta)}_{\rho_r}(z_{\beta})$ with $\rho_r=\sum_{t=0}^{b}x_{t}-1$
has an erasure pattern that can be (completely) recovered with the code generated by
$[G^{(\beta)}_{x_b-1}(z_{\beta}),\hdots,G^{(\beta)}_0(z_{\beta})]$, set $a_1=1$ and apply Algorithm 1 to recover $v^{(\beta)}_{\rho_r}(z_{\beta})$ and to get\\ $u^{(\beta)}_{\rho_r-x_b+1}(z_{\beta}), \hdots, u^{(\beta)}_{\rho_r}(z_{\beta})$, which is outputted as\\ $u^{(\beta)}_{\rho_r-x_b+1+\chi_{r-1}}(z_{\beta}), \hdots, u^{(\beta)}_{\rho_r+\chi_{r-1}}(z_{\beta})$. Replace the corresponding $\ast$-symbols in $\hat{v}^{(\beta)}_{\rho_r}(z_{\beta})$ with the recovered symbols and go to 5.\\
If $\hat{v}^{(\beta)}_{\rho_r}(z_{\beta})$
has an erasure pattern that can not be (completely) recovered, set $x_b\rightarrow x_b-1$ and go to 4.\\
\textbf{4}: If $x_b\neq 0$, go to 3, if $x_b=0$, go to 7.\\
\textbf{5}: $b\rightarrow b+1$\\
\textbf{6}: Go back to 2.\\
\textbf{7}: If $\chi_r=\sum_{l=0}^{r}\rho_l+r+1<d_{\alpha}+1$, continue with 8, otherwise stop with success.\\
\textbf{8}: If there exists $y\in\mathbb N$ such that $v^{(\beta)}_{\rho_r}(z_{\beta})$ with $\rho_r=\sum_{t=0}^{b-1}x_{t}+\mu_{\alpha}+y$ has an erasure pattern that can be recovered with the code generated by $[G^{(\beta)}_{\mu_{\alpha}}(z_{\beta}),\hdots,G^{(\beta)}_0(z_{\beta})]$, take $y$ minimal with this property, set $a_1=1$ and proceed with 9. If it does not exist, proceed with 18.\\
\textbf{9}: Decode $v^{(\beta)}_{\rho_r}(z_{\beta})$ in the code generated by $[G^{(\beta)}_{\mu_{\alpha}}(z_{\beta}),\hdots,G^{(\beta)}_0(z_{\beta})]$ 
to obtain $u^{(\beta)}_{\rho_r}(z_{\beta}), \hdots, u^{(\beta)}_{\rho_r-\mu_{\alpha}}(z_{\beta})$ applying Algorithm 1 and output it as\\ $u^{(\beta)}_{\rho_r+\chi_{r-1}}(z_{\beta}), \hdots, u^{(\beta)}_{\rho_r-\mu_{\alpha}+\chi_{r-1}}(z_{\beta})$.
\\
\textbf{10}: If there exists $1\leq w_c\leq \mu_{\alpha}$ such that $v^{(\beta)}_{\rho_r+\sum_{s=0}^{c}w_s-(c+1)(\mu_{\alpha}+1)}(z_{\beta})$ has an erasure pattern that could be recovered with the code generated by $[G^{(\beta)}_{\min (\mu_{\alpha},\mu_{\alpha}+y+\sum_{s=0}^{c}w_s-(c+1)(\mu_{\alpha}+1))}(z_{\beta}),\hdots, G^{(\beta)}_{w_c}(z_{\beta})]$, take $w_c$ minimal with this property, do the recovery applying Algorithm 1, replace the corresponding $\ast$-symbols with the recovered symbols and proceed with 11. If it does not exist, proceed with 18.\\
\textbf{11}: If $\mu_{\alpha}+y+\sum_{s=0}^{c}w_s-(c+1)(\mu_{\alpha}+1)\leq\mu_{\alpha}$, proceed with 14, otherwise proceed with 12.\\
\textbf{12}: $c\rightarrow c+1$\\
\textbf{13}: Go back to 10.\\
\textbf{14}: If $\chi_r=\sum_{l=0}^{r}\rho_l+r+1<d_{\alpha}+1$, continue with 15, otherwise stop with success.\\
\textbf{15}: For $m=0,\hdots,\rho_r$ set\\ $v_m^{(\beta)}(z_{\beta})=v_m^{(\beta)}(z_{\beta})-[G_{m-\rho_r}^{(\beta)}(z_{\beta}),\hdots, G_m^{(\beta)}(z_{\beta})]\begin{pmatrix}
u_{\rho_r}^{(\beta)}(z_{\beta})\\ \vdots\\ u_0^{(\beta)}(z_{\beta})
\end{pmatrix}$ and afterwards $\hat{v}(z_1,z_2)=\sum_{\rho_r+1\leq i\leq d_{\alpha}, 0\leq j\leq d_{\beta}}\hat{v}_{ij}z_{\alpha}^{i-\rho_r-1}z_{\beta}^{j}$,\\
$v(z_1,z_2)=\sum_{\rho_r+1\leq i\leq d_{\alpha}, 0\leq j\leq d_{\beta}}v_{ij}z_{\alpha}^{i-\rho_r-1}z_{\beta}^{j}$
and\\ $u(z_1,z_2)=\sum_{\rho_r+1\leq i\leq d_{\alpha}, 0\leq j\leq d_{\beta}}u_{ij}z_{\alpha}^{i-\rho_r-1}z_{\beta}^{j}$.\\
\textbf{16}: $r\rightarrow r+1$\\
\textbf{17}: Go to 1.\\
\textbf{18}: End of Algorithm 2.2., go back to main algorithm\\

This algorithm could be explained as follows:\\
Step \textbf{0} and \textbf{1} are initializations. In steps \textbf{2} to \textbf{4}, $x_0$ is choosen as large as possible such that $\hat{v}_{x_0-1}$ can be recovered, i.e. one tries to recover as many of the vectors $u^{(\beta)}_0(z_{\beta}), u^{(\beta)}_1(z_{\beta}),\hdots$ as possible in one step. Hereby, the condition $x_0\leq \mu_{\alpha}+1$ ensures that $u^{(\beta)}_0(z_{\beta})$ is among the recovered vectors, i.e. we have no gap at the beginning. We set $a_1=1$ if something could be corrected to pass the information to the main algorithm that we updated $\hat{v}$. One proceeds with $x_1,x_2,\hdots$ as long as possible, where the condition $x_b\leq \mu_{\alpha}+1$ ensures, that there is no gap between the $u^{(\beta)}_i(z_{\beta})$ that were recovered using $x_{b-1}$ and the $u^{(\beta)}_i(z_{\beta})$ that are recovered using $x_b$.\\
If it is not possible to recover everything with steps \textbf{1} to \textbf{7} of the algorithm, one proceeds with step \textbf{8}. In choosing such a $y$, we could decode further erasures but there will be a gap between the $u^{(\beta)}_i(z_{\beta})$ that were recovered using steps \textbf{1} to \textbf{7} and the $u^{(\beta)}_i(z_{\beta})$ recovered in step \textbf{9}. More precisely, we do not recover $u^{(\beta)}_{\sum_{t=0}^{b}x_t}(z_{\beta}),\hdots, u^{(\beta)}_{\sum_{t=0}^{b}x_t+y-1}(z_{\beta})$. In steps \textbf{10} to \textbf{13}, this gap is closed if this is possible. We start to recover the $u^{(\beta)}_i(z_{\beta})$ in the gap with largest indices $i$. Thereby, we choose $w_0$ as small as possible such that one of the already recovered vectors, here $u_{\rho_r-\mu_{\alpha}}^{(\beta)}(z_{\beta})$, is involved. This is necessary for the recovery to be possible since $y$ was chosen minimal and hence we know that recovery without any information from previous steps is not possible. Similiarly, one chooses all $w_c$ minimal such that recovery is possible which means that at least the vector with largest index has been already recovered using $w_{c-1}$. The decoding can be done in the code with generator matrix $[G^{(\beta)}_{\min (\mu_{\alpha},\mu_{\alpha}+y+\sum_{s=0}^{c}w_s-(c+1)(\mu_{\alpha}+1))}(z_{\beta}),\hdots, G^{(\beta)}_{w_c}(z_{\beta})]$ since the first $w_c$ vectors $u^{(\beta)}_i(z_{\beta})$ have been already recovered using $y, w_0,\hdots, w_{c-1}$ and maybe some of the last vectors already by using $x_0,\hdots,x_b$.

 After closing the gap, we check if everything is already recovered and if not we shift the indices of the vectors $\hat{v}^{(\beta)}_i(z_{\beta})$, $v^{(\beta)}_i(z_{\beta})$ and $u^{(\beta)}_i(z_{\beta})$ such that the first unrecovered part has now index zero (see step \textbf{15}) and start again at the beginning of the algorithm to recover the remaining erasures. For the following outputs, we have to consider that this shift has been done.\\

%


Like Algorithm 2.1 also Algorithm 2.2 is part of the main Algorithm 2 and depending on the erasure pattern it could be sufficient for recovering all erasures, as the following theorem states.

\begin{theorem}\ \\
If $\hat{v}(z_1,z_2)$ has an erasure pattern such that there are $\alpha, \beta \in\{1,2\}$ with $\alpha\neq\beta$ and indices $0\leq j_0<\cdots<j_l=\mu_{\alpha}+m_{\alpha}$ with $j_0\leq\mu_{\alpha}$ and $j_k-(j_{k-1}+1)\leq\mu_{\alpha}$ for $k=1,\hdots,l$ such that $v^{(\beta)}_{j_0}(z_{\beta})$ can be decoded in $[G^{(\beta)}_{j_0}(z_{\beta}),\hdots, G^{(\beta)}_{0}(z_{\beta})]$ and $v^{(\beta)}_{j_k}(z_{\beta})$ can be decoded in $[G^{(\beta)}_{j_k-(j_{k-1}+1)}(z_{\beta}),\hdots, G_0^{(\beta)}(z_{\beta})]$ for $k=1,\hdots,l$, then Algorithm 2.2 leads to complete recovery of all erasures.
\end{theorem}

\begin{remark}\ \\
The conditions of the preceding theorem are not necessary as they only give the conditions for being able to decode only with steps \textbf{1} to \textbf{7} of Algorithm 2.2.
\end{remark}

If $m_{\alpha}$ is known, everything that was said in the Case 1 (under the assumption that $m_{\alpha}$ is known) also applies here.

\subsection{Partial recovery of blocks in one direction}

Algorithm 2.1 and Algorithm 2.2 search for vectors $\hat{v}_i^{(\beta)}(z_{\beta})$ for which complete recovery is possible and if they find one, they do the corresponding recovery. However, if no complete recovery of $\hat{v}_i^{(\beta)}(z_{\beta})$ is possible, one can still use Algorithm 1 to recover as many erasures as possible in $\hat{v}_i^{(\beta)}(z_{\beta})$. Doing this, we do not optimize the order of recovery for the vectors $\hat{v}^{(\beta)}_{c}(z_{\beta})$ but go just straightforward because with partial recovery, we just recover some components of $v$ but cannot compute components of $u$. This idea is implemented in the following algorithm, which is also part of the main algorithm described later.\\

\textbf{Decoding algorithm 2.3}\\
\textbf{1}: Set $c=0$.\\
\textbf{2}: Recover as many erasure as possible in
$\hat{v}^{(\beta)}_{c}(z_{\beta})$
with the code generated by
$[G^{(\beta)}_{c}(z_{\beta}),\hdots,G^{(\beta)}_0(z_{\beta})]$, using Algorithm 1. If at least one erasure can be corrected, set $a_2=1$ and replace the corresponding $\ast$-symbols in $\hat{v}^{(\beta)}_{c}(z_{\beta})$ with the recovered symbols.\\
\textbf{3}: $c\rightarrow c+1$\\
\textbf{4}: If $c\leq e_{\beta}$, go to 2, if $c=e_{\beta}+1$, go to 5.\\
\textbf{5}: If $\mu_{\alpha}>e_{\beta}$, go back to main algorithm, otherwise go to 6.\\
\textbf{6}: Recover as many erasure as possible in
$\hat{v}^{(\beta)}_{c}(z_{\beta})$
with the code generated by
$[G^{(\beta)}_{\mu_{\alpha}}(z_{\beta}),\hdots,G^{(\beta)}_0(z_{\beta})]$, using Algorithm 1. If at least one erasure can be corrected, set $a_2=1$ and replace the corresponding $\ast$-symbols in $\hat{v}^{(\beta)}_{c}(z_{\beta})$ with the recovered symbols.\\
\textbf{7}: $c\rightarrow c+1$\\
\textbf{8}: If $c\leq d_{\alpha}$, go to 6, otherwise go back to main algorithm.\\

\subsection{Main algorithm - combining the recovery in both directions}

In this subsection, we put the parts of the previous subsections together to derive our main algorithm.

If one has recovered as much erasures as possible in the vectors\\ $\hat{v}_0^{(\beta)}(z_{\beta}),\hdots, \hat{v}^{(\beta)}_{d_{\alpha}}(z_{\beta})$ but there are still unrecovered erasures, one could switch $\alpha$ and $\beta$, i.e. the roles of $z_1$ and $z_2$.
Then, depending if $\mu_{\alpha}>\epsilon_{\beta}$ or not, we apply Algorithm 2.1 or Algorithm 2.2, respectively. 
If there are still erasures left after that, one could switch the variables again to be in the same situation as in the beginning of the decoding but with less erasures, which might enable the recovery of even more erasures. We will give a simple example to see how this switching of the variables could make it possible to recover more erasures than applying only Algorithm 2.1 or Algorithm 2.2.\\

\pagebreak

\textbf{Example 1}:\\
Assume that $v^{(2)}_0(z_2)=v_{00}+v_{01}z_2+\cdots$ arrives completely except that $v_{00}$ is erased completely and could not be recovered by trying to decode $v_1^{(2)}(z_2),v_2^{(2)}(z_2),\hdots$. Assume further that $v_1^{(1)}(z_1)=v_{01}+v_{11}z_1+v_{21}z_1^2+\cdots$ could be completely recovered, i.e. one obtains (amongst others) $u_0^{(1)}(z_1)$ and therefore, $v_0^{(1)}(z_1)=v_{00}+v_{10}z_1+\cdots$ is known. Hence, one knows $v_{00}$ and therefore, $v_0^{(2)}(z_2)$ is now recovered, which was not possible before.\\

In that way, one switches the variables until all erasures are recovered or one reaches the point that in both directions, with respect to $z_1$ and with respect to $z_2$, no further recovery is possible. Hence, the complete decoding procedure is performed according to the following algorithm.\\

\textbf{Decoding Algorithm 2}:\\
\textbf{1}: Set $\alpha=2$, $\beta=1$\\
\textbf{2}: If $\mu_{\alpha}>e_{\beta}$, apply Algorithm 2.1. If $\mu_{\alpha}\leq e_{\beta}$, apply Algorithm 2.2.\\
\textbf{3}: $\alpha\leftrightarrow\beta$\\
\textbf{4}: $a_1=0$.\\
\textbf{5}: If $\mu_{\alpha}>e_{\beta}$, apply Algorithm 2.1. If $\mu_{\alpha}\leq e_{\beta}$, apply Algorithm 2.2.\\
\textbf{6}: $\alpha\leftrightarrow\beta$.\\
\textbf{7}: If $a_1=0$, go to 8, if $a_1\neq 0$, go to 4.\\
\textbf{8}: $a_2=0$, $\alpha\leftrightarrow\beta$\\
\textbf{9}: Apply Algorithm 2.3\\
\textbf{10}: $\alpha\leftrightarrow\beta$\\
\textbf{11}: If $a_2=0$, go to 12, if $a_2\neq 0$, go to 4.\\
\textbf{12}: Apply Algorithm 2.3\\
\textbf{13}: $\alpha\leftrightarrow\beta$\\
\textbf{14}: If $a_2=0$, exit (no further recovery possible), if $a_2\neq 0$, go to 4.\\

In steps \textbf{2} to \textbf{5} we do the decoding procedure once with respect to every variable. In step \textbf{7}, we check if we recovered any new symbols in step \textbf{5}. If this is the case, we can continue with full recovery, otherwise we have to switch to the algorithm for partial recovery. In step \textbf{11}, we check if we recovered any new symbols with this algorithm in step \textbf{9}. If this is the case, we can go back to the algorithm for full recovery, otherwise we try partial recovery with respect ot the other variable. 
In the following, we illustrate Algorithm 2 with the help of an example.\\

\textbf{Example 2}:\\
 Assume that $G^{(1)}_0(z_1)$ and $G^{(1)}_1(z_1)$ are generator matrices of $(3,1,1)$ MDP convolutional codes and $[G_0^{(1)}(z_1)\ G^{(1)}_1(z_1)]$ is the generator matrix of a $(3,2,2)$ MDP convolutional code. Moreover, assume that $G^{(2)}_0(z_2)$ is the generator matrix of a $(3,1,1)$ MDP convolutional code and $[G_0^{(2)}(z_2)\ G^{(2)}_1(z_2)]$ is full column rank. We will give constructions for 2D convolutional codes with these properties later in this paper; see Construction 2 and Remark \ref{ex}.\\
A $(3,1,1)$ MDP convolutional code can recover at most $2$ out of $3$ erasures in windows of size $3$ or $6$, a $(3,2,2)$ MDP convolutional code can recover at most $1$ out of $3$ erasures in windows of size $3$, $6$, $9$ or $12$, where all these windows have to start with a window of size $3$ that contains erasures and where the preceding $\nu$ windows of size $3$ are free of erasures. Moreover, we have $\mu_{\alpha}=\epsilon_{\beta}=1$ for $\alpha,\beta\in\{1,2\}$.

Assume that the codeword (containing erasures) has the form $v(z_1,z_2)=\sum_{0\leq i,j\leq 4}v_{ij}z_1^{i}z_2^{j}$ and the following erasure pattern, where $\ast$ denotes an erasure and $v_{ij}=\begin{pmatrix}v_{ij,1}\\ v_{ij,2}\end{pmatrix}$ for $i,j=1,\hdots,4$.\\

\bigskip

\begin{tabular}{|c| c| c|c| c| c|}
\hline
$\hat{v}_{ij}$ & $j=0$ &  $j=1$ & $j=2$ &  $j=3$ & $j=4$  \\
\hline
$i=0$ & $\ast$  & $v_{01,1}$ & $\ast$  & $\ast$ & $\ast$ \\
 & $\ast$ & $v_{01,2}$ & $\ast$ & $\ast$  &   $\ast$ \\
  & $\ast$ & $v_{01,3}$ & $\ast$ & $v_{03,3}$  &   $\ast$ \\
\hline
$i=1$ & $\ast$  & $\ast$  & $v_{12,1}$ &  $\ast$  & $v_{14,1}$ \\
&  $\ast$ & $\ast$ & $v_{12,2}$ &  $\ast$ & $v_{14,2}$\\
&  $\ast$ & $\ast$ & $v_{12,3}$ &  $v_{13,3}$ & $v_{14,3}$\\
\hline
$i=2$ &  $v_{20,1}$ &  $\ast$ & $\ast$ & $\ast$ & $v_{24,1}$ \\
 & $v_{20,2}$ & $\ast$ &  $\ast$ &  $\ast$ &  $v_{24,2}$ \\
 & $v_{20,3}$ & $v_{21,3}$ &  $\ast$ & $v_{23,3}$ &  $v_{24,3}$ \\
\hline
$i=3$ & $\ast$ & $v_{31,1}$ & $\ast$ & $\ast$ &  $v_{34,1}$\\
 & $\ast$ & $v_{31,2}$ & $\ast$ & $\ast$ &  $v_{34,2}$\\
 & $\ast$ & $v_{31,3}$ &  $\ast$ &  $v_{33,3}$ & $v_{34,3}$\\
\hline
$i=4$ & $\ast$ &  $v_{41,1}$ &  $\ast$  & $\ast$ & $v_{44,1}$ \\
&  $\ast$ & $v_{41,2}$ &  $\ast$ &  $\ast$ &  $v_{44,2}$\\
&  $\ast$ & $v_{41,3}$ &  $\ast$ &  $v_{43,3}$ &  $v_{44,3}$\\
\hline
\end{tabular}\\

\bigskip
According to Algorithm 2, the decoding is done in the following steps:

\begin{enumerate}
\item
We apply Algorithm 2.2 with $\alpha=2$ and $\beta=1$:

\item[1.1]
As $v_1^{(1)}(z_1)$ cannot be decoded in the code generated by $[G_0^{(1)}(z_1)\ G_1^{(1)}(z_1)]$ and $v_0^{(1)}(z_1)$ cannot be decoded in the code generated by $G_0^{(1)}(z_1)$, one has $x_0=0$.

\item[1.2]
As $v_2^{(1)}(z_1)$ and $v_3^{(1)}(z_1)$ cannot be decoded in the code generated by $[G_0^{(1)}(z_1)\ G_1^{(1)}(z_1)]$ but $v_4^{(1)}(z_1)$ can, one has $y=3$ and obtains $u_3^{(1)}(z_1)$ and $u_4^{(1)}(z_1)$.

\item[1.3]
One gets $w_0=1$ and decodes $v_3^{(1)}(z_1)$ with the code generated by $G_1^{(1)}(z_1)$. This yields the missing $u_2^{(1)}(z_1)$ but as $v_2^{(1)}(z_1)$ cannot be recovered with the code generated by $G_1^{(1)}(z_1)$, $u_0^{(1)}(z_1)$ and $u_1^{(1)}(z_1)$ remain unrecovered.

\item
We apply Algorithm 2.2 with $\alpha=2$ and $\beta=1$: 

\item[2.1]
As $v_1^{(2)}(z_2)$ cannot be recovered with an $(3,2,2)$ MDP convolutional code, it cannot be recovered with the code generated by $[G_0^{(2)}(z_2)\ G_1^{(2)}(z_2)]$ (we do not even have to compute its column distances to know that). However $v_0^{(2)}(z_2)$ can be recovered with the code generated by $G_0^{(2)}(z_2)$.

\item[2.2]

As $v_4^{(2)}(z_2)$, $v_3^{(2)}(z_2)$ and $v_2^{(2)}(z_2)$ cannot be recovered with the code generated by $[G_0^{(2)}(z_2)\ G_1^{(2)}(z_2)]$ and $v_1^{(2)}(z_2)$ cannot be recovered with the code generated by $G_0^{(2)}(z_2)$, we go back to the main algorithm.

\item
We apply Algorithm 2.2 with $\alpha=2$ and $\beta=1$ but does not lead to further recovery.

\item
We apply Algorithm 2.3 with $\alpha=2$ and $\beta=1$, which leads to the recovery of $v_{10}$.

\item
We apply Algorithm 2.2 with $\alpha=1$ and $\beta=2$, which leads to the recovery of $v_{11}$, i.e. $v_1^{(2)}(z_2)$ is now completely recovered.

\item
We apply Algorithm 2.2 with $\alpha=2$ and $\beta=1$:

Now $v_1^{(1)}(z_1)$ can be decoded in the code generated by $[G_0^{(1)}(z_1)\ G_1^{(1)}(z_1)]$ (i.e. one has $x_0=2)$ and hence the missing $u_0^{(1)}(z_1)$ and $u_1^{(1)}(z_1)$ are obtained and thus, the whole information is recovered.

\end{enumerate}

\begin{remark}\ \\
If we had the same code and the same erasure pattern as in the previous example but $v_{24}$ would be erased, full recovery would not be possible (it would only be possible to recover $v_{00}$ and afterwards $v_{10}$ with Algorithm 2.3). However, assuming we know in addition that the degree of $u(z_1,z_2)$ with respect to $z_2$ is 3, i.e. $u_4^{(1)}(z_1)=0$, we could still decode $v_4^{(1)}(z_1)$ in the code generated by $G_1^{(1)}(z_1)$ and afterwards proceed as in the previous example achieving full recovery.
\end{remark}

\section{Construction of codes that are very well suited for the algorithm}\label{con}

\subsection{The case $n\leq 2k$}

In this case, we have $e_1=e_2=0$ and therefore, it is best if the codes generated by $G^{(1)}_0(z_1)$ and $G_0^{(2)}(z_2)$ are as good as possible. The best known convolutional codes over the erasure channel are the so-called complete MDP convolutional codes, which are a subclass of MDP convolutional codes and are defined as follows.

\begin{definition}\cite{vp}\label{com}
Let $H(z)=H_0+H_1z+\cdots H_{\nu}z^{\nu}\in\mathbb F[z]^{(n-k)\times n}$ be a parity-check matrix of the convolutional code $\mathcal{C}$ of rate $k/n$ and degree $\delta$. Set $L:=\lfloor\frac{\delta}{n-k}\rfloor+\lfloor\frac{\delta}{k}\rfloor$. Then
\begin{align}\label{ppc}
\mathfrak{H}:=\left(\begin{array}{ccccc}
H_{\nu} & \cdots & H_0 &   & 0 \\
  & \ddots &   & \ddots &   \\
0 &   & H_{\nu} & \cdots & H_0
\end{array}\right)  \in\mathbb F^{(L+1)(n-k)\times (\nu+L+1)n}
\end{align}
is called \textbf{partial parity-check matrix} of the code. Moreover, $\mathcal{C}$ is called \textbf{complete MDP} convolutional code if for any of its parity-check matrices $H(z)$, every full size minor of $\mathfrak{H}$ which is not trivially zero is nonzero.
\end{definition}

In addition to the erasure correcting capability of MDP convolutional codes, complete MDP convolutional codes admit the possibility to continue decoding if after a window with too many erasures one receives a window with a sufficiently low ratio of erasures; see \cite{vp} for more details.

\begin{theorem}\cite{cmdp}\ \\
Let $n,k,\delta\in\mathbb N$ with $k<n$ and $(n-k)\mid\delta$ and let $\gamma$ be a primitive element of a finite field $\mathbb F=\mathbb F_{p^N}$ with $N>(L+1)\cdot 2^{(\nu+2)n-k-1}$. Then $H(z)=\sum_{i=0}^{\nu}H_iz^i$ with $$H_i=\left[\begin{array}{ccc}
\gamma^{2^{in}} & \hdots & \gamma^{2^{(i+1)n-1}} \\
\vdots &  & \vdots \\
\gamma^{2^{(i+1)n-k-1}} & \hdots & \gamma^{2^{(i+2)n-k-2}}
\end{array}\right]\ \text{for}\ i=0,\hdots,\nu=\frac{\delta}{n-k}$$ is the parity-check matrix of an $(n,k,\delta)$ complete MDP convolutional code.
\end{theorem}

We use the preceding theorem to obtain an optimal construction for 2D convolutional codes with rate at least $1/2$.\\

\textbf{Construction 1}\\
Let $H_0^{(2)}(z_2)$ and $H_0^{(1)}(z_1)$ be parity-check matrices of the codes generated by $G_0^{(2)}(z_2)$ and $G_0^{(1)}(z_1)$, respectively. We set $H_0^{(2)}(z_2)=H(z_2)$ and $H_0^{(1)}(z_1)=H(z_1)$ where $H(z)$ should be defined as in the preceding theorem and $\nu$ could be chosen arbitrarily (the construction then automatically implies that the degree $\delta_0$ of the codes generated by $G_0^{(2)}(z_2)$ and $G_0^{(1)}(z_1)$ is $\delta_0=\nu(n-k)$). \\

This construction is optimal for the decoding algorithm in the case $e_1=e_2=0$ since in this case the algorithm only employs the codes generated by $G_0^{(2)}(z_2)$ and $G_0^{(1)}(z_1)$, which are complete MDP convolutional codes and therefore optimal. Moreover, the construction is also quite suitable for the decoding algorithm in the case $e_1\neq 0$ or $e_2\neq 0$ as the algorithm for this case can be considered as an extension of the algorithm for $e_1=e_2=0$.\\

Algorithm 2 together with Construction 1 should be illustrated with the help of the following example.\\

\textbf{Example 3}:\\
Assume that we do not know $m_1$ and $m_2$ and use the construction of the preceding theorem to obtain a 2D convolutional code, where $G_0^{(2)}(z_2)$ and $G_0^{(1)}(z_1)$ generate $(2,1,2)$ complete MDP convolutional codes (in \cite{j} the authors constructed such codes over the field of minimal possible size, which is much smaller than the one of the general construction we presented here). Such a code could recover all erasure patterns where in each sliding window of length 10 there are at most 5 erasures (see Theorem \ref{mdpp}). Moreover, if there is a window of length 14 with at most 5 erasures and there are not too many at the beginning and at the end of this window, then complete recovery of all symbols in this window is possible no matter how many erasures are outside this window (see \cite{vp}).\\

Assume that the received message (containing erasures) has the form $\hat{v}(z_1,z_2)=\sum_{0\leq i,j\leq 6}\hat{v}_{ij}z_1^{i}z_2^{j}$ with the following erasure pattern, where $\ast$ denotes an erasure and $v_{ij}=\begin{pmatrix}v_{ij,1}\\ v_{ij,2}\end{pmatrix}$ for $i,j=1,\hdots,6$.\\

\begin{tabular}{|c| c| c|c| c| c| c| c|}
\hline
$\hat{v}_{ij}$ & $j=0$ &  $j=1$ & $j=2$ &  $j=3$ & $j=4$ & $j=5$ & $j=6$ \\
\hline
$i=0$ & $v_{00,1}$  & $\ast$ & $v_{02,1}$  & $\ast$ & $v_{04,1}$ &  $v_{05,1}$  & $v_{06,1}$ \\
 & $v_{00,2}$ & $\ast$ & $v_{02,2}$ & $\ast$  &   $\ast$ & $v_{05,2}$ & $v_{06,2}$\\
\hline
$i=1$ & $v_{10,1}$  &  $\ast$ & $\ast$ & $\ast$ &  $\ast$ & $v_{15,1}$  & $v_{16,1}$ \\
&  $v_{10,2}$ & $\ast$ & $\ast$ & $\ast$ &  $\ast$ & $v_{15,2}$ & $v_{16,2}$\\
\hline
$i=2$ &  $\ast$ & $v_{21,1}$ & $v_{22,1}$ &  $\ast$ & $\ast$ & $v_{25,1}$ &  $v_{26,1}$ \\
 & $\ast$ & $v_{21,2}$ & $v_{22,2}$ & $\ast$ &  $\ast$ &  $v_{25,2}$ &  $v_{26,2}$\\
\hline
$i=3$ & $\ast$ & $\ast$ & $\ast$ & $\ast$ & $v_{34,1}$  & $v_{35,1}$  & $\ast$\\
 & $\ast$ & $\ast$ & $\ast$ & $\ast$ &  $v_{34,2}$ &  $v_{35,2}$ &  $\ast$\\
\hline
$i=4$ & $v_{40,1}$ &  $v_{41,1}$ &  $v_{42,1}$  &  $\ast$ &  $\ast$ & $v_{45,1}$ & $v_{46,1}$ \\
&  $v_{40,2}$ & $v_{41,2}$ &  $v_{42,2}$ & $\ast$ &  $\ast$ &  $v_{45,2}$ &  $v_{46,2}$\\
\hline
$i=5$ & $v_{50,1}$  & $v_{51,1}$ &  $v_{52,1}$  & $\ast$ &  $\ast$ & $v_{55,1}$  & $v_{56,1}$ \\
 &  $v_{50,2}$ &  $v_{51,2}$ &  $v_{52,2}$ & $\ast$ &  $\ast$  & $v_{55,2}$ & $v_{56,2}$\\
\hline
$i=6$ & $v_{60,1}$ & $v_{61,1}$ & $v_{62,1}$ &  $\ast$ &  $\ast$ & $v_{65,1}$ &  $v_{66,1}$ \\
 &  $v_{60,2}$  & $v_{61,2}$ & $v_{62,2}$  & $\ast$ &  $\ast$ & $v_{65,2}$ & $v_{66,2}$\\
\hline
\end{tabular}\\

\normalsize
\bigskip
We start applying Algorithm 2 with $\alpha=1$ and $\beta=2$ (i.e. this time we start the decoding in the other direction, which makes no difference):

\begin{enumerate}
\item
The erasure pattern of $\hat{v}_0^{(2)}(z_2)=\hat{v}_{00}+\cdots+\hat{v}_{06}z_2^6$ allows recovery with the complete MDP convolutional code generated by $G_0^{(2)}(z_2)$. Hence, one obtains $v_{01}$, $v_{03}$ and $v_{04,2}$.
\item
Next consider $\hat{v}_1^{(2)}(z_2)$. It contains two many erasures to be corrected. Hence, we change the roles of $z_1$ and $z_2$.
\item
The erasure pattern of $\hat{v}_0^{(1)}(z_1)$ allows recovery (see Theorem \ref{mdpp}) and we obtain $v_{20}$ and $v_{30}$.
\item
Since we already recovered $v_{01}$ in part 1 of this example, the erasure pattern of $\hat{v}_1^{(1)}(z_1)$ allows recovery and we obtain $v_{11}$ and $v_{31}$.
\item
Continue with $\hat{v}_2^{(1)}(z_1)$, which can be recovered and we obtain $v_{12}$ and $v_{32}$.
\item
As $v_3^{(1)}(z_1)$ is completely erased and thus, cannot be recovered, we switch the roles of $z_1$ and $z_2$ again.
\item
Since we have recovered $v_{11}$ and $v_{12}$ in part 4 and part 5, respectively, decoding of $\hat{v}_1^{(2)}(z_2)$ is now possible and we obtain $v_{13}$ and $v_{14}$.
\item
The erasure pattern of $\hat{v}_2^{(2)}(z_2)$ allows recovery and we obtain $v_{23}$ and $v_{24}$.
\item
As we already recovered $v_{30}$, $v_{31}$ and $v_{32}$ in part 3, part 4 and part 5, respectively, it is possible to recover $v_3^{(2)}(z_2)$, i.e. to obtain $v_{33}$ and $v_{36}$.
\item
The remaining erasures in $\hat{v}_4^{(2)}(z_2)$, $\hat{v}_5^{(2)}(z_2)$ and $\hat{v}_6^{(2)}(z_2)$ can be recovered.
\end{enumerate}


If we take the erasure pattern of the preceding example but assume that $v_{22}$ is erased, we would fail in part 5 of this example to recover $v_2^{(1)}(z_1)$. As the recovery of $v_{11}$ in part 4 is not enough to enable the decoding of $\hat{v}_1^{(2)}(z_2)$, Algorithm 2 is not able to recover all the erasures of this pattern (also Algorithm 2.3 does not lead to any recovery).\\
However if we assume that $m_1$ and $m_2$ are known, Algorithm 2 would still be able to decode this erasure pattern. It recovers the vectors $v_i^{(\beta)}(z_{\beta})$ in the following order:\\
$v_0^{(2)}(z_2)$, $v_6^{(2)}(z_2)$, $v_5^{(2)}(z_2)$, $v_4^{(2)}(z_2)$, $v_0^{(1)}(z_1)$, $v_1^{(1)}(z_1)$, $v_6^{(1)}(z_1)$, $v_5^{(1)}(z_1)$, $v_4^{(1)}(z_1)$, $v_1^{(2)}(z_2)$, $v_2^{(2)}(z_2)$, $v_3^{(2)}(z_2)$\\

\subsection{The case $n>2k$}
We choose $e_1=e_2=\lceil\frac{n}{k}\rceil-2$, i.e. maximal such that $k(e_i+1)<n$.\\
In order to extend Construction 1, which ensures that $G_0^{(2)}(z_2)$ and $G_0^{(1)}(z_1)$ generate complete MDP convolutional codes, to the case that $e_i\neq 0$, it would be necessary to construct the not yet determined coefficient matrices of $G(z_1,z_2)$ in such way that $[G_0^{(2)}(z_2),\hdots, G^{(2)}_{e_2}(z_2)]$ and $[G_0^{(1)}(z_1),\hdots,G_{e_1}^{(1)}(z_1)]$ generate convolutional codes.
If $H_0^{(2)}(z_2)$ and $H_0^{(1)}(z_1)$ are fixed as in Construction 1, i.e. $G_0^{(2)}(z_2)$ and $G_0^{(2)}(z_1)$ are fixed, too, it is possible to choose the remaining coefficients such that $[G_0^{(1)}(z_1),\hdots,G^{(1)}_{e_1}(z_1)]$ has full column rank. However, one does not know anything about the decoding properties of $[G_0^{(1)}(z_1),\hdots,G_{l}^{(1)}(z_1)]$ for $1\leq l\leq e_1$. Moreover, one does not know if $[G_0^{(2)}(z_2),\hdots,G^{(2)}_r(z_2)]$ for $1\leq r\leq e_2$ generate convolutional codes, i.e. have full column rank, which is a necessary condition for our algorithm.\\
A first idea would be to use the construction for complete MDP convolutional codes that we used for Construction 1 to obtain that $[G_0^{(1)}(z_1),\hdots, G^{(1)}_{e_1}(z_2)]$ generates a complete MDP convolutional code. But again, this construction does not imply that $[G_0^{(1)}(z_1),\hdots,G_l^{(1)}(z_1)]$ for $1\leq l\leq e_1$ are MDP convolutional codes nor that $[G_0^{(2)}(z_2),\hdots, G^{(2)}_r(z_2)]$ for $1\leq r\leq e_2$ even generate convolutional codes. For example, if we use the construction for complete MDP convolutional codes that we applied for Construction 1 to construct $[G^{(1)}_0\ G^{(1)}_1](z_1)$ in such way that it generates an $(3,2,1)$ complete MDP convolutional code, we get $[G^{(1)}_0\ G^{(1)}_1](z_1)=\left[\begin{array}{cc}\gamma^3-\gamma^{13}+(\gamma^{17}-\gamma^{20})z_2 & \gamma^{12}-\gamma^5\\ \gamma^{12}-\gamma^2 & \gamma^{11}-\gamma\\ (\gamma^8-\gamma^5)z_2 & 1-\gamma^3\end{array}\right]$. However, with this construction, $G_0^{(1)}(z_1)$ does not generate an MDP convolutional code, which could be easily seen from the fact that its last entry has no constant coefficient.\\
To overcome this problem, we propose the following construction, which uses the following proposition.\\

\begin{proposition}\cite{dr16}
Let $\gamma$ be a primitive element of a finite field $\mathbb F=\mathbb F_{p^N}$ and $B=[b_{i,l}]$ be a matrix over $\mathbb F$ with the following properties
\begin{enumerate}
\item if $b_{i,l}\neq 0$, then $b_{i,l}=\gamma^{\beta_{i,l}}$ for a positive integer $\beta_{i,l}$
\item if $b_{i,l}=0$, then $b_{i',l}=0$ for any $i'>i$ or $b_{i,l'}=0$ for any $l'<l$
\item if $l<l'$, $b_{i,l}\neq 0$ and $b_{i,l'}\neq 0$, then $2\beta_{i,l}\leq\beta_{i,l'}$
\item if $i<i'$, $b_{i,l}\neq 0$ and $b_{i',l}\neq 0$, then $2\beta_{i,l}\leq\beta_{i',l}$.
\end{enumerate}
Suppose $N$ is greater than any exponent of $\gamma$ appearing as a nontrivial term of any minor of $B$. Then $B$ is superregular, i.e. all minors that are not trivially zero are nonzero.
\end{proposition}

\medskip

\textbf{Construction 2}:\\
Set $\hat{k}=(\mu_2+1)k$ and $\mathfrak{G}^{(1)}(z_1):=[G_0^{(1)}(z_1),\hdots,G^{(1)}_{\mu_2}(z_1)]:=\sum_{j=0}^{\mu_2}\mathfrak{G}_jz_1^j$. Define $\mathfrak{G}_j:=\left[\begin{array}{ccc}
\gamma^{2^{jn}} & \hdots & \gamma^{2^{jn+\hat{k}-1}} \\
\vdots &  & \vdots \\
\gamma^{2^{(j+1)n-1}} & \hdots & \gamma^{2^{(j+1)n+\hat{k}-2}}
\end{array}\right]$ for $j=0,\hdots,\mu_1$.
Then, one has the following properties:

(1) $[G_k^{(1)}(z_1),\hdots, G_l^{(1)}(z_1)]$ for $0\leq k\leq l\leq \mu_2$ generate MDP convolutional codes (see Theorem \ref{cd} and the preceding proposition).

(2) $[G_{00}, G_{10},\hdots,G_{\mu_1 0},G_{01},G_{11},\hdots,G_{i1},\hdots,G_{0\mu_2},\hdots,G_{\mu_1\mu_2}]$ has nonzero fullsize minors.


(3) $G_{\mu_1\mu_2}$ is of full rank, i.e. $m_{\alpha}=deg_{z_{\alpha}}v(z_1,z_2)+\mu_{\alpha}\leq deg_{z_{\alpha}}\hat{v}(z_1,z_2)+\mu_{\alpha}$ for ${\alpha}\in\{1,2\}$.\\

The second property implies that $[G_0^{(2)}(z_2),\hdots,G_{e_2}^{(2)}(z_2)]_{z_2=0}$ has full column rank and hence $[G_0^{(2)}(z_2),\hdots,G_{e_2}^{(2)}(z_2)]$ has full column rank.\\
From the third property, one gets the information that $u_i(z_{\beta})=0$ for $i>d_{\alpha}+\mu_{\alpha}$, which simplifies the decoding (see the description of advantages if $m_{\alpha}$ is known).

\begin{remark}\label{ex}\ \\
As used in Example 2, for $k=1$, $n=3$ and $\mu_1=\mu_2=1$, using Construction 2, it holds that $G_0^{(2)}(z_2)$ is the generator matrix of an MDP convolutional code. This follows from Theorem \ref{cd} since for $G_0^{(2)}(z_2)$, we have $G^c_1=\left(\begin{array}{cc} \gamma & 0\\ \gamma^2 & 0\\ \gamma^4 & 0\\ \gamma^2 & \gamma\\ \gamma^4 & \gamma^2\\ \gamma^8 & \gamma^4 \end{array}\right)$.
\end{remark}

\section{Performance evaluation}

In this section, we want to consider the number of erasures that could be corrected in a square of a certain size assuming that there are no erasures around this square. For this, we assume that $G_0^{(2)}(z_2)$ and $G_0^{(1)}(z_1)$ generate an $(n,k,\delta_1)$ and an $(n,k,\delta_2)$ MDP convolutional code, respectively. Note that Construction 1 has this property for $\delta_1=\delta_2$.

\begin{theorem}\ \\
If $G_0^{(2)}(z_2)$ and $G_0^{(1)}(z_1)$ generate an $(n,k,\delta_1)$ and an $(n,k,\delta_2)$ MDP convolutional code, respectively, and in a square of size $(L_1+1)n\times (L_2+1)n$ there are not more than $(L_1+L_2+2)(n-k)-(n-1)$ erasures, where $L_1:=\left\lfloor\frac{\delta_1}{k}\right\rfloor+\left\lfloor\frac{\delta_1}{n-k}\right\rfloor$ and $L_2:=\left\lfloor\frac{\delta_2}{k}\right\rfloor+\left\lfloor\frac{\delta_2}{n-k}\right\rfloor$, our algorithm can correct all erasures inside this square no matter where they are located.
\end{theorem}

\begin{proof}
Assume that it is not possible to recover all erasures with Algorithm 2. Moreover, assume that we were able to correct all erasures in the coefficient vectors of $v_0^{(2)}(z_2),\hdots,v_i^{(2)}(z_2)$ and $v^{(1)}_0(z_1),\hdots,v^{(1)}_j(z_1)$ for some $i\in\{0,\hdots,L_2\}$ and some $j\in\{0,\hdots,L_1\}$ but we are neither able to correct $v^{(2)}_{i+1}(z_2)$ nor $v^{(1)}_{j+1}(z_1)$ (using the MDP codes generated by $G_0^{(2)}(z_2)$ or $G_0^{(1)}(z_1)$, respectively). This means that there are at least $(L_2+1)(n-k)+1$ erasures in the coefficient vectors of $v^{(2)}_{i+1}(z_2)$ and at least $(L_1+1)(n-k)+1$ erasures in the coefficient vectors of $v^{(1)}_{j+1}(z_1)$, see Theorem \ref{mdpp}. As the only common coefficient vector of these two polynomials is $v_{i+1,j+1}$, at most $n$ of these erasures are identical, which leads to at least $(L_1+L_2+2)(n-k)-(n-2)$ erasures.
\end{proof}

\begin{remark}\ \\
There is an erasure pattern of $(L_1+L_2+2)(n-k)-(n-2)$ erasures in a square of size $(L_1+1)n\times (L_2+1)n$ that cannot be corrected with our algorithm if $G_0^{(2)}(z_2)$ and $G_0^{(1)}(z_1)$ generate MDP but not reverse MDP convolutional codes.\\
\end{remark}

\begin{proof}
(1) The erasure pattern is as follows: $v_{00}$ is erased completely, the first $(L_2+1)(n-k)-(n-1)$ symbols of $v_{01},\hdots,v_{0,L_2}$ are erased and the first $(L_1+1)(n-k)-(n-1)$ symbols of $v_{10},\hdots,v_{L_1,0}$ are erased.
\end{proof}

In many applications, erasures have the tendency to occur in bursts. Especially for these bursts of erasures the use of 2D convolutional codes together with the decoding algorithms of this paper is very advantageous. This is due to the fact that a burst of erasures in one direction can be recovered by decoding in the other direction.

\begin{theorem}
Let us assume that we receive 2-dimensional data in the order $v_{00},\hdots,v_{0,\deg_{z_1}(v)},v_{10},\hdots $, i.e. at first the first line, then the second line and so on (see the following table).\\
All algorithms of this paper (together with the corresponding construction) can correct a burst of erasures of length $(L_1+1)(n-k)\cdot\deg_{z_1}(v(z_1,z_2))$.
\end{theorem}

\begin{tabular}{|c| c| c|c|}
\hline
$\hat{v}_{ij}$ & $j=0$ &  $\cdots$ & $j=\deg_{z_1}(v(z_1,z_2))$ \\
\hline
$i=0$ &  $\ast$  & $\cdots$ & $\ast$  \\
\hline
$\vdots$ & $\vdots$ &  & $\vdots$ \\
\hline
$i=(L_1+1)(n-k)$ &  $\ast$ &  $\cdots$ &  $\ast$  \\
\hline
$i=(L_1+1)(n-k)+1$ & $v_{(L_1+1)(n-k)+1,0}$ & $\cdots$ & $v_{(L_1+1)(n-k)+1,\deg_{z_1}(v(z_1,z_2))}$\\
\hline
$\vdots$ & $\vdots$ &   &  $\vdots$   \\
\hline
$i=(L_1+1)n$ & $v_{(L_1+1)n,0}$ &  $\cdots$ &  $v_{(L_1+1)n,\deg_{z_1}(v(z_1,z_2))}$  \\
\hline
\end{tabular}\\

\begin{proof}
The preceding erasure pattern could be recovered decoding\\ $v_0^{(1)}(z_1),\hdots,v_{\deg_{z_1}(v)}^{(1)}(z_1)$ in the MDP convolutional code generated by $G_0^{(1)}(z_1)$.
\end{proof}

If one does not know which patterns of erasures the matrices\\ $[G_{m}(z_i),\hdots,G_l(z_i)]$ for $m,l\in\{0,\hdots,\mu_{\alpha}\}$ with $m\leq l$ for $i\in\{1,2\}$ could correct (which is only true if we do not know the structure of the matrix $G(z_1,z_2)$), then another possibility of decoding would be just to start to try to decode $v_0(z_2)$, then $v_1(z_2)$ and so on as long as possible. If one reaches a point where this is not possible anymore, one could proceed with the decoding of $v_0(z_1)$, $v_1(z_1)$ and so on and recover in each step as much as possible. One could continue this and switch variables after each run (with the same parameters) until no further recovery is possible or everything is recovered.

However, if one has e.g. MDP codes and knows which erasure patterns can be corrected, one can decrease the effort and speed up the decoding by the algorithms described in this paper. This can also be seen with the help of Example 2 as there one can skip $v_0(z_2)$ and start with the decoding of $v_1(z_2)$.\\

Our main advantage over \cite{dec} is that there the vectors $v_i(z_{\beta})$ are considered isolated of each other. With the method of \cite{dec}, it is enough if $v_0(z_2)$ and $v_0(z_1)$ cannot be decoded in the codes with parity-check matrices $H_0(z_2)$ and $H_0(z_1)$, respectively, to let the whole decoding fail.

\section{Conclusion}
In this paper, we presented the first complete decoding algorithm for 2D convolutional codes over the erasure channel. Moreover, we provided constructions of 2D convolutional codes targeted to this algorithm. An interesting problem for future research is to develop also a decoding algorithm for 2D convolutional codes that allows correction of transmission errors.

\section*{Acknowledgements}
This work is supported by The Center for Research and Development in
Mathematics and Applications (CIDMA) through the Portuguese Foundation
for Science and Technology (FCT - Funda\c{c}\~ao para a Ci\^encia e a Tecnologia), references UIDB/04106/2020 and UIDP/04106/2020, by the Swiss National Science Foundation grant n. 188430 and the German Research Foundation grant LI 3101/1-1.

\bibliography{mybibfile}

\end{document}